\documentclass[aps,prb,preprint,showpacs]{revtex4}
\usepackage{bm}
\usepackage{epsfig}
\usepackage{color}
\usepackage{amssymb}
\usepackage{amsmath}
\usepackage{wrapfig}
\usepackage[center]{caption}
\usepackage[utf8]{inputenc}

\begin{document}

\title{
	Nanograin ferromagnets from non-magnetic bulk materials: the case of
	gold nanoclusters
}
\author{N\'ora~Kucska and Zsolt~Gul\'acsi}
\affiliation{Department of Theoretical Physics, University of
Debrecen, H-4010 Debrecen, Bem ter 18/B, Hungary}

\begin{abstract} 

The ferromagnetism of Au nanograins is analysed based on a two-dimensional 
itinerant lattice model with on-site Coulomb repulsion, many-body spin-orbit 
interactions, and holding two hybridized bands, one correlated and one 
uncorrelated. Using periodic boundary conditions in both directions, an exact 
ferromagnetic ground state is deduced for this non-integrable system by 
applying special techniques based on positive semidefinite operators.

\end{abstract}


\maketitle

\section{Introduction}

\subsection{About Nanomagnetism}
Nanomagnetism attracts nowadays extreme attention given by both the theoretical challenges and the broad technological application possibilities of the subject \cite{1}. In this field we often encounter situations when magnetic moments present in bulk magnets produce magnetism also at nanosize level, as e.g. in the cases of magnetic nanocomposites \cite{2} or nanomagnets containing magnetic ions \cite{3}, whose properties are usually connected to the behaviour of magnetic domains or domain walls in constrained geometry. But the field of nanomagnetism presents also examples where materials not containing magnetic moments possessing bulk behaviour placed far from magnetism, present magnetic properties at nanoscale. These materials trigger a special chapter of nanomagnetism, in which the magnetic properties are intimately connected to the nanosize of the sample, and this is the subject which attracted our interest and attention during the research that leads to this article. A typical example on this line is the case of gold, which does not contain magnetic atoms, is diamagnetic in macroworld, but becomes ferromagnetic around 2 nm size \cite{4}. We are concentrating below to this property in our aim to explain it in the light of the presently existing experimental data. 

\subsection{Nanomagnetism of gold nanoparticles}

It is known that none of the 4d or 5d elements is magnetic in bulk form i.e. does not possess in macroscopic form non-zero magnetic moment in zero applied external field \cite{5}. But since 1999 it is known that Au nanoparticles become ferromagnetic \cite{6}. The surprising observation is real since it has been checked in several laboratories \cite{4,7,8}, see also \cite{5}. When prepared, the nanoparticles are in most cases defended (ligand-coated, functionalized) at the surface \cite{9}, and this fact leads to several theories aiming to explain the source of this type of nanomagnetismn by the ligand present on the surface (as e.g. \cite{10,11,12,13}). But the ferromagnetism appears also without functionalization \cite{7}, remains when one removes from the surface the ligand \cite{7,14}, and it was observed that photochemically prepared nanoparticles which never have contacted any added ligand have similar magnetic properties \cite{14}. Consequently, not the ligand on the surface causes the magnetic properties, so explaining this type of nanomagnetism other reasons must be used for explanation. One underlines here, that the ligand could modify the magnetic moment (as even observed e.g. \cite{15}), but the source of the magnetization is completely different. We further note that the presence of magnetic impurities (as e.g. Fe, Ni, Co, Mn, etc.) which could cause the appearance of magnetism experimentally can be excluded \cite{5,16}. 
Turning back to theory, one often encounters mostly mean-field type of descriptions (e.g. \cite{17}) which takes into account spin-spin type of interactions without clarifying precisely why, how and from where the magnetic moments come from, and if are present, why should we neglect the correlations (absent in mean-field). In essence one particle description (as e.g. \cite{18}), similar to mean-field treatments, also lack inter-particle correlations. The DFT descriptions [as e.g. \cite{19} using spin-polarized generalized gradient approximation (GGA)] has as well shortcomings in correctly reproducing the many-body correlations \cite{20}, and finally leads to ferromagnetism considering the nanocluster an icosahedral superatom in which magnetic alignment is caused by a “superatom Hunds rule”. This last is deduced from two-particle exchange, whose application in realistic many-body case is questionable (see e.g. \cite{5}). We underline here the many-body correlations since the current literature strongly stresses that the correlation effects in Au are important \cite{21,22,23}.
Properly described correlation effects in describing gold nanoparticles have been considered in \cite{24}. Here a two-dimensional lattice is considered with itinerant spin-1/2 carriers and periodic boundary conditions in both directions as describing a closed surface. Note that for nanograins more than the half of atoms are disposed on the surface, and the itinerant charged carriers, given by the Coulomb repulsion are also disposed on the surface. In describing the inter-particle correlations on-site Coulomb repulsion has been taken into account in the many-body Hamiltonian, for which an exact ground state solution has been deduced. For small samples (e.g. $LxL, L=12$ lattice) the ground state turns out to be ferromagnetic. The result does not require rigorous spherical surface: it demands closed surface, (even small) local Coulomb repulsion and many-body quantum mechanics. The deduced solution emerges in the small size limit, and for $L\ \rightarrow\infty$ loses its importance. That is why by increasing the size, the total magnetization decreases, and at $L\ \gg\ 1$ gradually disappears as observed experimentally \cite{5}. Since the model is non-integrable (one band 2D Hubbard model) a special technique based on positive semidefinite operator properties has been used in order to deduce the exact ground state. The ferromagnetic property arises from the parallel alignment of spins necessary to avoid the Hubbard (on-site Coulomb) repulsion in order to minimize the ground state energy. The fact that the ground state is a coherent quantum many-body state is supported by experimental data. Indeed, if one introduces impurities in the system, the quantum coherence is diminished, hence the magnetic moment decreases as observed by introducing Fe impurities in Au nanograins \cite{25}.

\section{Placing the model closer to the real system}

The model presented in \cite{24} shows a potential possibility, but in fact is a toy model that should be pushed more closely to the real analysed system. First, the published studies show that the multiband character of Au is important to be taken into consideration. Indeed, it was observed (e.g. \cite{22}) that in gold, the 5d orbital has a prominent contribution to the conduction electron system. The presence of 5d orbitals accentuate that during the description, at least a correlated band, and a non-correlated band  should be taken into account, and since direct evidence of 5d magnetism is not present \cite{22}, a hybridized two band system should represent the starting point. At the level of interactions, since we are confined to short distances, the on-site Coulomb repulsion U must represent the starting point, which is able to describe properly the correlation effects in the system.

 But the published results show that also another interaction, namely the many-body spin-orbit coupling (SOC), plays a major role. This was observed already at the level of one particle descriptions \cite{18}, during the study of the importance of the correlation effects \cite{23}, but nowadays, given by the observation that relativistic aspects must be included in the proper description \cite{23}, the importance of the SOC becomes to be a clear fact \cite{26,27,28,29}. We note that in fact, the large SOC at Au surface is known for many years now \cite{30,31}.

\begin{figure}
	\begin{center}
		\includegraphics[width=0.35\textwidth]{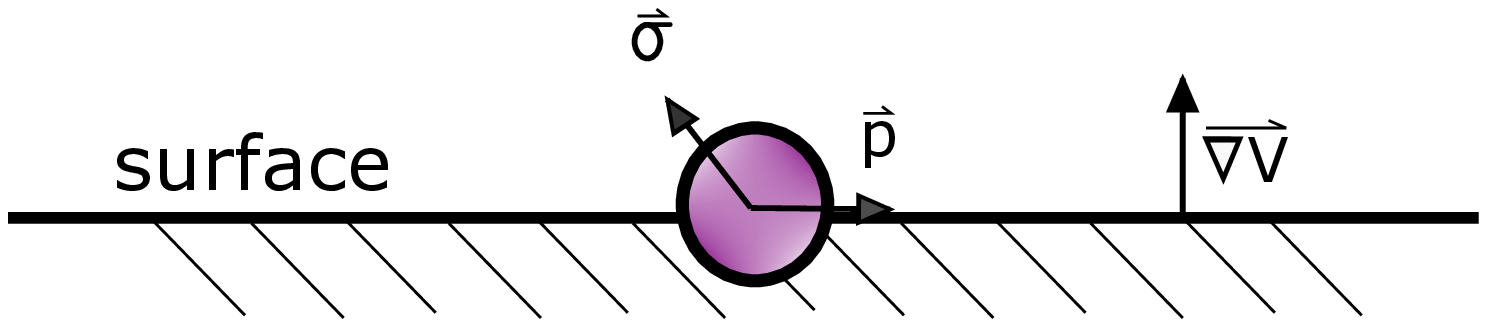}
	\end{center}
$\lambda\vec{\sigma}\cdot (\vec{\nabla}V \times\vec{p}) $
	\captionsetup{justification=centering}
	\caption{The many-body spin-orbit coupling}
\end{figure}

On its turn, the many-body spin-orbit interaction is in fact a relativistic effect (relativistic correction to the Schrödinger equation) which shows that if a carrier holding spin $(\vec{\sigma})$ is moving (i.e. has momentum p), and during this movement it feels a potential gradient $\vec{\nabla} V$ not colinear to$\ \vec{p}$, an interaction of the form ${\hat{H}}_{SO}=\lambda\vec{\sigma}\cdot\left(\vec{\nabla}V \times\vec{p}\right)$ appears, which represents the many-body spin orbit interaction, $\lambda$ being its coupling-constant (i.e. strength), (see Fig.1).

 Since in central field $\vec{\nabla} V \sim\vec{r}$, and $\vec{r}\times\vec{p}$ provides the orbital moment, ${\hat{H}}_{SO}$ indeed describes the SOC. But more importantly, at a surface (in the present case at the nanograin surface), perpendicular to the surface automatically a $\vec{\nabla} V$ potential gradient appears, so ${\hat{H}}_{SO}$ will be automatically present.

  The $\lambda$ value is typically small, of order ${10}^{-3}$ eV, but in the Au case, one knows \cite{30} that it attains in order 0.1 eV. That is why SOC is important for gold nanograins as well. We must underline that $\lambda\ll\ U$ usually holds, but even in this case $\lambda$ produces essential effects because breaks the double spin-projection degeneracy of each band. This is the reason why the perturbative treatment of ${\hat{H}}_{SO}$ in presence of $U\ >\ 0$ gives erroneous results, explaining why we use exact methods for description. We further note that at second quantized level, the presence of SOC gives rise in fact in the system Hamiltonian to spin-flip type of hopping terms of specific form \cite{32,33}.  In our opinion, the main effect of the ligand on the surface is to influence the ${\hat{H}}_{SO}$ strength.
 
 Our aim in this paper is to show that the main conclusions of \cite{24} concerning ferromagnetism in gold nanograins remain true et exact level also in the case of two-band treatment, and presence of ${\hat{H}}_{SO}$ in the many-body 2D Hamiltonian containing the on-site Coulomb repulsion in the correlated band.

\section{The used model Hamiltonian and its results}

\subsection{The positive semidefinite form of the Hamiltonian}

The Hamiltonian of the system in its original form has the expression  
\begin{equation}
\hat{H}= \sum_{p,p'} \sum_{i,r} \sum_{\sigma,\sigma'} (t_{i,i+r}^{p,p';\sigma,\sigma'} \hat{c}^{\dagger}_{p,i,\sigma} \hat{c}_{p',i+r,\sigma'}+ H.c.) +\sum_i U \hat{n}_{d,i,\uparrow} \hat{n}_{d,i,\downarrow}
\end{equation} 	 
where the first term represents the kinetic $\left( {\hat{H}}_{kin} \right)$, while the second the interaction part $\left({\hat{H}}_{int}\right)$,  of the Hamiltonian (on-site Coulomb repulsion $U\ >\ 0$ in the correlated d-band), and $p,p^\prime=s,d$ are representing the band indices where $ s$ symbolycally represents the non-correlated band (being here concretely of sp type). 
Please note that only the d-band is 
correlated in Eq.(1), the s-band being considered without Coulomb repulsion
is uncorrelated.
The sum over $i$ incorporates all lattice sites, and periodic boundary conditions are taken into account in both directions. In ${\hat{H}}_{kin}$ the $p\neq p^\prime$ contributions are representing hybridization terms. Denoting by $x_1,x_2$ the Bravais vectors of the 2D system, $r=x_1,x_2$ describes nearest neighbour hoppings, while $r=0$ characterizes on-site potentials. The SOC is taken into account by the Rashba term \cite{34} i.e.

\begin{eqnarray}
t_{i,i+x_1}^{p,p;\uparrow,\downarrow}=-t_{i,i+x_1}^{p,p;\downarrow,\uparrow}=t_p^R,\ \ t_{i,i+x_2}^{p,p;\uparrow,\downarrow}=t_{i,i+x_2}^{p,p;\downarrow,\uparrow}=-it_p^R. 
\end{eqnarray}

\begin{figure}
	\begin{center}
		\includegraphics[width=0.28\textwidth]{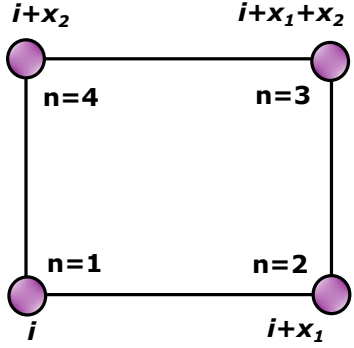}
	\end{center}
	\captionsetup{justification=centering}
	\caption{The plaquette connected to the site $i$ 
used for defining the block operators $B_{m,i}$ (see text above
Eq.(3)). The Bravais vectors are denoted by $x_1,x_2$ . The in-plaquette 
notation of lattice sites is provided by $n=1,2,3,4$.}
\end{figure}

The starting $\hat{H}$ in (1) incorporates all desired characteristics underlined in the previous chapter: surface many-body quantum physics (i.e. 2D character), two-bands character (i.e. $p=s,d$), presence of correlations (the on-site Coulomb repulsion $U$ in the correlated d band), and spin-orbit coupling (Rashba term). For a proper treatment of the common influence of the spin-orbit coupling and correlations, we deduce exact ground states on the line of \cite{32}, but using only the Rashba term in SOC. Since the system is non-integrable, a special technique is used based on positive semidefinite operator properties which has been tested in detail in several circumstances \cite{35,36,37}, being checked as well in the case of nanostructures \cite{38}, multiband structure \cite{39}, specific correlation effects \cite{40}, effects of the confinement \cite{41}, even disordered systems \cite{42}, detailed review papers being also available relating the technique \cite{43,44}. The technique starts with the transformation of the starting Hamiltonian presented in (1) in positive semidefinite form. For this reason we take elementary plaquettes defined on each lattice site i, see Fig.2, one plaquette containing four sites, namely $i,i+x_1,i+x_2,i+x_1+x_2$, representing the four corner sites of an unit cell. The in-plaquette notation of this four sites is given by the index $n=1,2,3,4$ in positive trigonometric circulation direction $( i_1=i,i_2=,i+x_1,i_3=i+x_1+x_2,i_4=i+x_2)$.

After this step, on each plaquette we introduce two block operators $m=1,2$ as ${\hat{B}}_{1,i}=\hat{V}\left(1\right)_i, {\hat{B}}_{2,i}=\hat{V}\left(2\right)_i$, which represent a linear combination of fermionic operators ${\hat{c}}_{p,i,\sigma} (p=s,d)$ acting on each site of the elementary plaquette on both bands: 

\begin{eqnarray}
\hat{V}\left(1\right)_i=\sum_{p,n,\sigma}{v\left(1\right)_{p,n,\sigma}{\hat{c}}_{p,i_n,\sigma}},\hat{V}\left(2\right)_i=\sum_{p,n,\sigma}{v\left(2\right)_{p,n,\sigma}{\hat{c}}_{p,i_n,\sigma}},
\end{eqnarray}

where $p=s,d$, one has $n=1,2,3,4$ and $\sigma=\uparrow,\downarrow$. The numerical prefactors $v\left(1\right)_{p,n,\sigma}$ and $v\left(2\right)_{p,n,\sigma}$ in (3) are the block operator parameters whose value is deduced in the transformation process. Note that the block operators contain both spin indices in order to be able to reproduce in the positive semidefinite ${\hat{B}}_{m,i} {\hat{B}}_{m,i}^{\dagger}$ expressions the spin-flip hopping terms introduced in the Hamiltonian by the spin-orbit contributions. We further underline that two block operators per plaquette are needed in order to exclude from the positive semidefinite form the contributions not present in the starting Hamiltonian (1). Concerning the strategy  we use, we remember that, if $\hat{O}$ is a positive semidefinite operator, than for all its arbitrary matrix elements the relation $\langle u|\hat{O}\ |u\rangle \geq 0$ holds, hence all its eigenvalues are non-negative. This is the reason why transforming a Hamiltonian in a positive semidefinite form, and deducing its eigenvector for the minimum eigenvalue (zero), we are able to find the ground state of the system, even if the system is non-integrable (i.e. the number of constants of motion is much less than the number of degrees of freedom, which is the case for almost all many body systems in nature).

Using the introduced block operators, the starting Hamiltonian in (1) is exactly transformed in the positive semidefinite form 
\begin{equation}
\hat{H}=\hat{P}+C\ ,\hat{P}={\hat{P}}_B+{\hat{P}}_U.
\end{equation}

Here the first positive semidefinite contribution is defined as ${\hat{P}}_B=\sum_{i}\sum_{m=1,2}{{\hat{B}}_{m,i}{\hat{B}}_{m,i}^\dag}$ while the second positive semidefinite contribution is given by$ {\hat{P}}_U=U\sum_{i}{\hat{P}}_i$ where $U\ >\ 0$ holds, and ${\hat{P}}_i={\hat{n}}_{i,\uparrow}^d{\hat{n}}_{i,\downarrow}^d-\left({\hat{n}}_{i,\uparrow}^d+{\hat{n}}_{i,\downarrow}^d\right)+1$, while $C$ is a scalar. Note that ${\hat{P}}_i$ is a positive semidefinite operator since attains its minimum eigenvalue zero, when there is at least one electron present at the lattice site $i$. We underline that given by the structure of the Hamiltonian presented in (4), and the fact that $\hat{P}$ is a positive semidefinite operator, the ground state $\left|\Psi_g\right\rangle$ of the system is given by the relation

\begin{eqnarray}
\hat{P}\ \left|\Psi_g\right\rangle\ =0, 	
\end{eqnarray}
while the eigenvalue connected to $\left|\Psi_g\right\rangle$  (i.e. the ground state energy $E_g$) is given by the relation $E_g=C$. As underlined previously, (5) provides the exact ground state independent on dimensionality and integrability.

\subsection{The matching equations}
 It can be observed that the starting Hamiltonian in (1) depends on the coupling constants of the Hamiltonian (as $t_{i,i+r}^{p,p^\prime;\sigma,\sigma^\prime}$, $U$, where $r=0,x_1,x_2; p,p\prime=s,d;$ and $\sigma,\sigma^\prime=\uparrow,\downarrow$ ), while the transformed Hamiltonian in (4) depends on the block operator parameters (as $v\left(m\right)_{p,n,\sigma}$, where $m=1,2;\ p=s,d;\ n=1,2,3,4;$ and $\sigma=\uparrow,\downarrow$). This means that since the $\hat{H}$ in (1) and (4) is exactly the same, interdependences must be present in between the Hamiltonian coupling constants and block operator parameters. These relations are called to be the matching equations. 

The matching equations are obtained by a) explicitly writing all different contributions in (1) and (4) -- note that for this the products and sums present in ${\hat{P}}_B$ and ${\hat{P}}_U$ in (4) must be explicitly calculated and written -, and b) equating the coefficients of all different operators from (1) to the coefficient of the same operator in (4).

For example, let us take from the starting Hamiltonian in (1) the $\uparrow,\downarrow$ spin-flip hopping terms in the $p=s$ band which has the expression ${\hat{h}}_1=t_{i,i+x_1}^{s,s,\uparrow,\downarrow}{\hat{c}}_{s,i,\uparrow}^\dag{\hat{c}}_{s,i+x_1,\downarrow}$ . We observe that such contribution in (4) emerges in ${\hat{P}}_D$, namely (if one fixes the m=1 index value, hence one considers only the ${\hat{B}}_{1,i}$ operators) in two products, namely ${\hat{I}}_1={\hat{B}}_{1,i}{\hat{B}}_{1,i}^\dag$ and ${\hat{I}}_2={\hat{B}}_{1,j}{\hat{B}}_{1,j}^\dag$ where the lattice site j is placed in the lattice just below the site i. Indeed, from  ${\hat{I}}_1$ we obtain the term ${\hat{b}}_{1,i}=v\left(1\right)_{s,2,\downarrow}v\left(1\right)_{s,1,\uparrow}^\ast{\hat{c}}_{s,i+x_1,\downarrow}{\hat{c}}_{s,i,\uparrow}^\dag$, while from ${\hat{I}}_2$ the contribution ${\hat{b}}_{1,j}=v\left(1\right)_{s,3,\downarrow}v\left(1\right)_{s,4,\uparrow}^\ast{\hat{c}}_{s,i+x_1,\downarrow}{\hat{c}}_{s,i,\uparrow}^\dag$ (note that $j+x_1+x_2=i+x_1, and j+x_2=i$). One knows that ${\hat{b}}_{1,i}$ can be also written as ${\hat{b}}_{1,i}=-v\left(1\right)_{s,2,\downarrow}v\left(1\right)_{s,1,\uparrow}^\ast{\hat{c}}_{s,i,\uparrow}^\dag{\hat{c}}_{s,i+x_1,\downarrow}$, and similarly, ${\hat{b}}_{1,j}=-v\left(1\right)_{s,3,\downarrow}v\left(1\right)_{s,4,\uparrow}^\ast{\hat{c}}_{s,i,\uparrow}^\dag{\hat{c}}_{s,i+x_1,\downarrow}$, hence one reobtains exactly the operator expression from ${\hat{h}}_1$. Other operators of the form ${\hat{h}}_1$ are not present in (4) if $m=1$ holds, but similar two contributions (${\hat{b}}_{2,i}$ and ${\hat{b}}_{2,j}$) one finds in the $m=2$ case. Consequently, since from the exact transformation of (1) to (4) (between others), also ${\hat{h}}_1 =\sum_{m=1,2}  (\hat{b}_{m,i} +{\hat{b}}_{m,j} )$ must hold, the matching equation connecting $t_{i,i+x_1}^{s,\uparrow,\downarrow}$ to block operator parameters becomes

\begin{equation}
-t_{i,i+x_1}^{s,s,\uparrow,\downarrow}=\sum_{m=1,2}\left(v\left(m\right)_{s,1,\uparrow}^\ast v\left(m\right)_{s,2,\downarrow}+v\left(m\right)_{s,4,\uparrow}^\ast v\left(m\right)_{s,3,\downarrow}\right).
\end{equation}
 
Similarly, let us take a next nearest-neighbor (unit cell diagonal) spin-flip hopping term in the same band $p=s$, namely ${\hat{h}}_2=t_{i,i+x_1+x_2}^{s,s,\uparrow,\downarrow}{\hat{c}}_{s,i,\uparrow}^\dag{\hat{c}}_{s,i+x_1+x_2,\downarrow}$.

One observes that since the starting Hamiltonian in (1) contains only nearest-neighbor hopping terms, ${\hat{h}}_2$ is missing from (1), consequently $t_{i,i+x_1+x_2}^{s,s,\uparrow,\downarrow}=0$ holds. But ${\hat{h}}_2$ type of contributions are present in (4), namely in ${\hat{P}}_D$, concretely (for the case of fixed $m=1$) in ${\hat{I}}_1={\hat{B}}_{1,i}{\hat{B}}_{1,i}^\dag$. Indeed, effectuating the product in ${\hat{I}}_1$, we find the contribution $ {\hat{d}}_{1,i}=v\left(1\right)_{s,3,\downarrow}v\left(1\right)_{s,1,\uparrow}^\ast{\hat{c}}_{s,i+x_1+x_2,\downarrow}{\hat{c}}_{s,i,\uparrow}^\dag=-v\left(1\right)^\ast s,1,\uparrow v\left(1\right)_{s,3,\downarrow}{\hat{c}}_{s,i,\uparrow}^\dag{\hat{c}}_{s,i+x_1+x_2,\downarrow}$ which (with another numerical prefactor) contains the operator ${\hat{h}}_2$ missing from the starting Hamiltonian (1). The same type of ${\hat{h}}_2$ contribution is obtained from (4) also at $m=2$ in the form ${\hat{d}}_{2,i}=v\left(2\right)_{s,3,\downarrow}v\left(2\right)_{s,1,\uparrow}^\ast{\hat{c}}_{s,i+x_1+x_2,\downarrow}{\hat{c}}_{s,i,\uparrow}^\dag=-v\left(2\right)^\ast s,1,\uparrow v\left(2\right)_{s,3,\downarrow}{\hat{c}}_{s,i,\uparrow}^\dag{\hat{c}}_{s,i+x_1+x_2,\downarrow}$. Consequently, because other ${\hat{h}}_2$ type of contributions are no more present in $\hat{P}$, and since for the exact transformation of (1) to (4), also the equality $0={\hat{h}}_2=\sum_{m=1,2}{\hat{d}}_{m,i}$ must hold, a matching equation emerges of the form

\begin{equation}
-t_{i,i+x_1+x_2}^{s,s,\uparrow,\downarrow}=\sum_{m=1,2}{v\left(m\right)_{s,1,\uparrow}^\ast}v\left(m\right)_{s,3,\downarrow}=v\left(1\right)_{s,1,\uparrow}^\ast v\left(1\right)_{s,3,\downarrow}+v\left(2\right)_{s,1,\uparrow}^\ast v\left(2\right)_{s,3,\downarrow}=0. 
\end{equation}

Equation (7) clearly shows why we need two ( $B_{1,i},B_{2,i}$ i.e. $m=1,2$) block operators for the transformation in the positive semidefinite form: the reason is that this helps us in eliminating those contributions in the transformed positive semidefinite Hamiltonian expression (4), which are not present in the starting Hamiltonian (1). Without two block operators used in the present case, expressions like (7) would nullify the uniquely introduced block operator, so would eliminate the possibility of the transformation of the Hamiltonian in positive semidefinite form. In a similar manner all matching equations can be deduced. In the present case their number is relatively high, namely 74, and are presented together with their solution in Appendix A. Writing the matching equations one finishes the first step of the method (the transformation of the Hamiltonian in positive semidefinite form). After this job, the second step of the technique follows, namely the solution of the matching equations. For this system of equations, the Hamiltonian coupling constants are considered known quantities, while the unknown variables are the block operator parameters. The difficulty of this second step is that the matching equations are representing a coupled, non-linear, complex algebraic system of equations containing (in realistic cases describing real materials) a relatively high number of components (and standard numerical softwares for coupled nonlinear system of equations with high number of  equations, in the general case, are not known). But, with our background and experience in solving such systems, the solutions can be usually deduced, and for the present case, are also presented in Appendix A.

\subsection{The deduced ground state and its physical properties} 

Now the third step of the technique follows: the deduction of the ground state $\left|\Psi_g\right\rangle$. This is done from (5), and one finds

\begin{equation}
|\Psi_g \rangle= \left[\prod_{i=1}^{N_\Lambda}\left(\prod_{m=1,2}{\hat{B}}_{m,i}^\dag\right){\hat{D}}_i^\dag \right] {\hat{R}}_{N_1}^\dag |0\rangle, 
\end{equation}
where $\left|0\right\rangle$ is the bare vacuum with no fermions present. One has ${\hat{D}}_i^\dag={\hat{c}}_{d,i,\uparrow}^\dag+{\hat{c}}_{d,i,\downarrow}^\dag$, while  ${\hat{R}}_{N_1>0}^\dag=\prod_{\alpha=1}^{N_1}{\hat{c}}_{p_\alpha,k_\alpha,\sigma_\alpha}^\dag$ and ${\hat{R}}_{N_1=0}^\dag=1$, where $1\le N_1<N_\Lambda$ holds, ${\hat{c}}_{p,k,\sigma}$ being the Fourier transform of ${\hat{c}}_{p,i,\sigma},p_\alpha$ is arbitrarily s or d; $\sigma_\alpha$ is arbitrarily $\uparrow,\downarrow; k_\alpha$ (with the condition that at $\alpha\neq\alpha^\prime$ one has $k_\alpha\neq k_{\alpha^\prime}$) is an arbitrary $k$ value from the first Brillouin zone. 

\begin{figure}
	\begin{center}
		\includegraphics[width=0.48\textwidth]{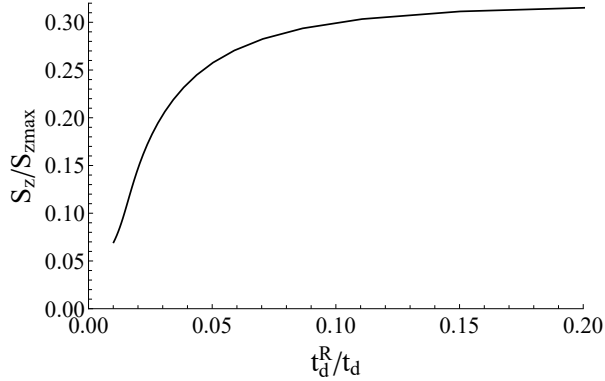}
	\end{center}
	\captionsetup{justification=centering}
	\caption{The ground state expectation value of the normalized $S_z$ total spin per number of particles, in function of the Rashba coupling of the correlated band ($t_d^R$) in units of the nearest neighbour hopping ($t_d$) in the same band}
\end{figure}

The uniqueness of the ground state can be demonstrated on the line of \cite{43}. For $N_1=0$,  $\left( N_1>0 \right)$,  the ground state (8) corresponds to $3/4$, (above $3/4$), system filling. The (5) is satisfied because $i$) ${\hat{B}}_{m,i}^\dag{\hat{B}}_{m,i}^\dag=0$, hence ${\hat{P}}_D\left|\Psi_g\right\rangle=0$ holds, and $ii$) one has on each site at least one carrier in $\left|\Psi_G\right\rangle$ hence also ${\hat{P}}_U\left|\Psi_g\right\rangle=0$ is fulfilled. Below one considers the $N_1=0$ case, but similar results one finds at $N_1>0$ as well.

As shown by Fig.3, the normalized ground state expectation value of the total spin z component per number of particles increases in function of the SOC coupling in the absence of external magnetic field, signalling ferromagnetic order. When the number of atoms in the volume exceeds considerably the number of atoms on the surface, the here deduced ground state loses its importance and the magnetic state disappears. 
Indeed, let us consider a toy exemplification of this aspect by considering
FCC cells (in which usually Au crystalizes) disposed in cubic grains whose
size is d. Let further introduce the ratio $f=N_s/N_t$, where $N_t$ is the 
total number of atoms from the system, and $N_s$ is the number of atoms on the 
surface. For 1 cell ($N_t=14$ atoms, d=4.14 $\AA$), f=100 $\%$. For 8 cells  
($N_t=63$, d=8.28 $\AA$) the studied ratio is still $f=80 \%$. But for $16^3$
= 4096 cells (17969 atoms, d=66.24 $\AA$ = 6.62 nm) one has only $f=17\%$. Our 
model being a 2D model, takes into account the surface, so the deduced ground 
state is the ground state of the Au grain only if f is high. For $f=17\%$ 
(i.e. $83\%$ of the atoms from the system are neglected), our deduced ground 
state is no more the ground state of the system, hence ferromagnetism 
dissapears. This surely happens when the grain mass density (for $16^3$ FCC 
cells here $\rho=20.21$ g/$cm^3$) approaches from above the bulk density 
(for gold, $\rho=19.3$ g/$cm^3$).

The ferromagnetic spin alignment is enforced by the on-site Coulomb repulsion (Hubbard U term) i.e. many-body correlations, since in this case the ground state energy attains its minimum value by avoiding double occupancy in the direct r-space, the k-space itinerant states providing components on all lattice sites. The here deduced ground state type \cite{32} emerges above system half filling, but similar ferromagnetic ground state can be deduced as well below system half filling \cite{33}. Consequently, doping is necessary for the ferromagnetism to occur as observed experimentally \cite{45}.

\section{Summary and conclusions}

After a survey of experimental data and used theoretical descriptions, the ferromagnetism of gold nanoparticles is explained by a deduced 2D exact many-body ground state determined by the common effect of the spin-orbit coupling, Coulomb correlations, hybridized bands and doping on a closed itinerant nanosurface. Besides to help the doping process, the role of the ligand on the surface turns out to influence only the strength of the spin-orbit coupling. A similar description probably works also in the case of Pd nanograins where the importance of similar effects have been pointed out \cite{46,47,48,49,50}.
We further note that if an arbitrary non-magnetic and metallic compound becomes 
ferromagnetic at nanoscales, and presents similar material properties as 
detailed for Au at the beginning of Section II, the here presented theory 
would explain this behavior.

\appendix

\section{The detailed system of matching equations and their solutions}

The matching equations In this section the detailed matching equations are 
presented. The first set of 16 equations are describing the nearest-neighbor 
hopping $\left(p=p^\prime\right)$ and hybridizations $\left(p\neq p^\prime\right)$
in the $x_1$ direction.

\begin{equation}
-t_{i,i+x_1}^{p,p^\prime,\sigma,\sigma^\prime}=\sum_{m=1,2}\left(v\left(m\right)_{p,1,\sigma}^\ast v\left(m\right)_{p^\prime,2,\sigma^\prime}+v\left(m\right)_{p,4,\sigma}^\ast v\left(m\right)_{p^\prime,3,\sigma^\prime}\right). 	
\end{equation}

Similarly, in the $x_2$ direction, the nearest-neighbor hopping and hybridizations are provided by the following 16 equations

\begin{equation}
-t_{i,i+x_2}^{p,p^\prime,\sigma,\sigma^\prime}=\sum_{m=1,2}\left(v\left(m\right)_{p,1,\sigma}^\ast v\left(m\right)_{p^\prime,4,\sigma^\prime}+v\left(m\right)_{p,2,\sigma}^\ast v\left(m\right)_{p^\prime,3,\sigma^\prime}\right).
\end{equation}

The next nearest-neighbor hoppings and hybridizations missing from the starting Hamiltonian provide homogenous equations, again 16 along the main diagonal of the unit cell i.e. $r=x_2+x_1$ (in the left side), and new 16 equations along the anti-diagonal of the unit cell i.e. $r=x_2-x_1$ (in the righ side)

\begin{equation}
0=\sum_{m=1,2}{v\left(m\right)_{p,1,\sigma}^\ast}v\left(m\right)_{p^\prime,3,\sigma^\prime},\ \ 0=\sum_{m=1,2}{v\left(m\right)_{p,2,\sigma}^\ast}v\left(m\right)_{p^\prime,4,\sigma^\prime}. 
\end{equation}

Up to this moment 64 equations have been presented connected to non-local Hamiltonian terms. One has further 10 local terms as follows: For the case of local hybridizations $\left(p\neq p^\prime\right)$, given by the relation $t_{i,i,r=0}^{p,p^\prime\neq p,\sigma,\sigma^\prime}=\left(t_{i,i,r=0}^{p^\prime\neq p,p,\sigma^\prime,\sigma}\right)^\ast$, one has 4 matching equations, namely

\begin{eqnarray}
-t_{i,i,r=0}^{p,p^\prime\neq p,\sigma,\sigma^\prime}=\sum_{m=1,2,}\sum_{n=1,2,3,4}{v\left(m\right)_{p,n,\sigma}^\ast}v\left(m\right)_{p^\prime,n,\sigma^\prime}. 
\end{eqnarray}

In the case of band-diagonal local terms 6 more matching equations are present. In the case of spin-diagonal such contributions, one has 4 matching equations, namely
\begin{eqnarray}
t_{i,i,r=0}^{p,p,\sigma,\sigma}+U\delta_{p,d}-\chi=\sum_{m=1,2,}\sum_{n=1,2,3,4}\left|v\left(m\right)_{p,n,\sigma}\right|^2, 	
\end{eqnarray}

where $\chi$ is a supplementary parameter determined by the matching equations and entering in the $C$ constant in the transformed Hamiltonian from (4). The spin non-diagonal $(\sigma\prime\ \neq\sigma)$ and $p^\prime=p$ components, given by the relation $t_{i,i,r=0}^{p,p,\sigma,\sigma^\prime}=\left(t_{i,i,r=0}^{p,p,\sigma^\prime,\sigma}\right)^\ast$, provide only two new matching equations, namely

\begin{equation}
t_{i,i,r=0}^{p,p,\uparrow,\downarrow}=\sum_{m=1,2,}\sum_{n=1,2,3,4}{v\left(m\right)_{p,n,\uparrow}^\ast v\left(m\right)_{p,n,\downarrow}}.
\end{equation}
 		
Besides, the C constant from the transformed Hamiltonian in (4) is given by 

\begin{eqnarray}
C=\chi N-U_dN_\Lambda-\sum_{i}\sum_{m=1,2}{\{{\hat{B}}_{m,i}}{\hat{B}}_{m,i}^\dag\},\  	
\end{eqnarray}
where $N$ represents the number of carriers,$N_\Lambda$ the number of lattice sites, while $\{\hat{X},\hat{Y}\}=\hat{X}\hat{Y}+\hat{Y}\hat{X}$ holds.

The solutions of the matching equations In order to solve the matching equations one starts with the homogenous equations with minimum number of components (as e.g. (A3)), going further by gradually using the more complicated equations. First one finds that all $v\left(1\right)_{p,n,\sigma}$ can be expressed in terms of $v\left(2\right)_{p,n,\sigma}$ parameters as 

\begin{eqnarray}
v\left(1\right)_{p,n,\sigma}=K_nv\left(2\right)_{p,n,\sigma}, 
\end{eqnarray}
where the proportionality constants $K_n$ are given by 

\begin{equation}
K_1=-\frac{1}{x}, K_2=-\frac{1}{y}, K_3=x^\ast, K_4=y^\ast,
\end{equation}
where $x,y$ are ($\neq0,\infty$) parameters.
Furthermore it results that from the remaining unknown $v\left(2\right)_{p,n,\sigma}$, those containing p=s band index can be expressed in terms of the p=d components as follows

\begin{eqnarray}
v\left(2\right)_{s,1,\sigma}=\frac{xu_\sigma}{y}v\left(2\right)_{d,3,-\sigma}^\ast,
v\left(2\right)_{s,2,\sigma}=u_\sigma v\left(2\right)_{d,4,-\sigma}^\ast,
\nonumber\\
v\left(2\right)_{s,3,\sigma}=-\frac{u_\sigma}{x^\ast y}v\left(2\right)_{d,1,-\sigma}^\ast,\ \ v\left(2\right)_{s,4,\sigma}=-\frac{u_\sigma}{\left|y\right|^2}v\left(2\right)_{d,2,-\sigma}^\ast,  
\end{eqnarray}
where $u_\sigma$ are new $(\neq0,\infty)$ parameters. 
After this stage several solution classes are present, from which we exemplify here the $u=u_\uparrow=-u_\downarrow$ case (this condition provides zero value for the local spin-flip contributions). In this situation all the remaining $v\left(2\right)_{d,n,\sigma}$ parameters can be expressed in terms of $v\left(2\right)_{d,1,\sigma}$ as follows 

\begin{eqnarray}
v\left(2\right)_{d,3,\uparrow}=-\frac{e^{i\phi_3}}{x^\ast}v\left(2\right)_{d,1,\uparrow},\ \ v\left(2\right)_{d,3,\downarrow}=\frac{e^{i\phi_3}}{x^\ast}v\left(2\right)_{d,1,\downarrow},
\nonumber\\
v\left(2\right)_{d,4,\uparrow}=-\frac{e^{i\phi_4}}{y^\ast}v\left(2\right)_{d,2,\uparrow}, v\left(2\right)_{d,4,\downarrow}=\frac{e^{i\phi_4}}{y^\ast}v\left(2\right)_{d,2,\uparrow}, 	
\end{eqnarray}
where $\phi_3,\phi_4$ are arbitrary phases. Then one finds

\begin{eqnarray}
v\left(2\right)_{d,2,\uparrow}\ =\frac{\sqrt2|y|e^{i\gamma}}{\left|\left(\left|y\right|-1\right)\right|}v\left(2\right)_{d,1,\downarrow},\ \ v\left(2\right)_{d,2,\downarrow}=\ \frac{\sqrt2|y|e^{-i\gamma}}{\left|\left(\left|y\right|-1\right)\right|}v\left(2\right)_{d,1,\uparrow},
\end{eqnarray}
where $\gamma$ is an arbitrary phase, furthermore, for $y=\left|y\right|e^{i\phi_y}$, one has $x=\left|x\right|e^{i\phi_y}$ where $\left|x\right|=\left(\left|y\right|-1\right)/\left(\left|y\right|+1\right)$ holds. 
The last two unknown block operator parameters are deduced from 

\begin{eqnarray}
(t_{i,i+x_1}^{s,d,\uparrow,\uparrow}e^{-i\eta}+t_{i,i+x_2}^{s,d,\uparrow,\uparrow})=A_\uparrow e^{i\delta_\uparrow} ,
 {A}_\uparrow=\left|t_{i,i+x_1}^{s,d,\uparrow,\uparrow}e^{-i\eta}+t_{i,i+x_2}^{s,d,\uparrow,\uparrow}\right|,	
 \nonumber\\
-\left(t_{i,i+x_1}^{s,d,\uparrow,\uparrow}e^{-i\eta}-t_{i,i+x_2}^{s,d,\uparrow,\uparrow}\right)=A_\downarrow e^{i\delta_\downarrow},{\ \ A}_\downarrow=\left|t_{i,i+x_1}^{s,d,\uparrow,\uparrow}e^{-i\eta}-t_{i,i+x_2}^{s,d,\uparrow,\uparrow}\right|, 
\end{eqnarray}
where $\eta=\phi_3+\phi_y$. Starting from this relation, introducing

\begin{equation}
b=\sqrt{\frac{\left(\left|y\right|-1\right)^2}{2\sqrt2(|y|^2+1)}\frac{|y|}{|u|}} ,	\ u=\left|u\right|e^{i\phi_u},   
\end{equation}     			    
where $\phi_u$ is an arbitrary phase, one has

\begin{equation}
v\left(2\right)_{d,1,\uparrow}=b\sqrt{A_\uparrow}e^{i\left(\gamma+\phi_u+\delta_\uparrow\right)/2},\ \ v\left(2\right)_{d,1,\downarrow}=b\sqrt{A_\downarrow}e^{i\left(-\gamma+\phi_u+\delta_\downarrow\right)/2}. 
\end{equation}
	     
Furthermore

\begin{eqnarray}
\frac{\left|y\right|}{\left|u\right|}=\frac{-t_{i,i+x_1}^{d,d,\uparrow,\uparrow}}{\sqrt{A_\uparrow A_\downarrow}\cos{\frac{\delta_\uparrow-\delta_\downarrow}{2}}},\ \ 
tang\ \frac{\delta_\uparrow-\delta_\downarrow}{2}=\frac{t_{i,i+x_2}^{d,d,\uparrow,\uparrow}}{t_{i,i+x_1}^{d,d,\uparrow,\uparrow}}\ , 	
\end{eqnarray}
and $\left|y\right|$ remains arbitrary. Note that the presented solution preserves $t_{i,j}^{p,p^\prime,\uparrow,\uparrow}=t_{i,j}^{p,p^\prime,\downarrow,\downarrow}$.

\end{document}